\documentclass[12pt]{article}

\font\SC=cmcsc10 scaled 1440

\def\rdots{\mathinner{\mkern1mu\raise1pt\vbox{\kern1pt\hbox{.}}\mkern2mu
   \raise4pt\hbox{.}\mkern2mu\raise7pt\hbox{.}\mkern1mu}}
\newcommand{\Z}{{\rm Z\kern-.35em Z}}
\newcommand{\bP}{{\rm I\kern-.15em P}}
\newcommand{\Q}{\kern.3em\rule{.07em}{.65em}\kern-.3em{\rm Q}}
\newcommand{\R}{{\rm I\kern-.15em R}}
\newcommand{\h}{{\rm I\kern-.15em H}}
\newcommand{\C}{\kern.3em\rule{.07em}{.55em}\kern-.3em{\rm C}}
\newcommand{\T}{{\rm T\kern-.35em T}}
\newcommand{\D}{{\kern-.5em /}}

\begin{document}

\openup 1.5\jot

\centerline{The Axial  Anomaly Revisited $^*$}

\vspace{1in}
\centerline{Paul Federbush}
\centerline{Department of Mathematics}
\centerline{University of Michigan}
\centerline{Ann Arbor, MI 48109-1109}
\centerline{(pfed@math.lsa.umich.edu)}
\vspace{4.0in}

$^*$ This work was supported in part by the National Science Foundation under Grant No. PHY-92-04824 .
\vfill\eject

$$ $$
$$ $$

\centerline{{\bf ABSTRACT}}

\vspace{1in}

\indent

We consider theories with  gauged chiral fermions in which there are abelian anomalies, and no nonabelian anomalies (but there may be nonabelian gauge fields present).  We construct an associated theory that is gauge-invariant, renormalizable, and with the same particle content, by adding a finite number of terms to the action.  Alternatively one can view the new theory as arising from the original theory by using another regularization, one that is gauge-invariant.  The situation is reminiscent of the mechanism of adding Fadeev-Popov ghosts to an unsatisfactory gauge theory, to arrive at the usual quantization procedure.  The models developed herein are much like the abelian Wess-Zumino model (an abelian effective theory with a Wess-Zumino counterterm), but unlike the W-Z model are renormalizable!  

Details of the approach are worked out explicitly for the special case of a single massless Dirac fermion, for which we couple one abelian gauge field to the vector current and another abelian gauge field to the axial current.

\vfill\eject

\section{Introduction.}

\indent
We consider a massless Dirac fermion coupled to two gauge bosons described by the formal Lagrangian 
\begin{equation}
{\cal L} = i \bar{\Psi} \gamma^\mu(\partial_\mu + ie A_\mu + i\gamma^5 fB_\mu)\Psi - {1 \over 4} \; F_{\mu \nu} F^{\mu \nu}  - {1 \over 4} \; G_{\mu \nu} G^{\mu \nu} \end{equation} 
with 
\begin{equation}   F_{\mu \nu} = \partial_\mu A_\nu - \partial_\nu A_\mu    \end{equation}
and        
\begin{equation}       G_{\mu \nu} = \partial_\mu B_\nu - \partial_\nu B_\mu.
\end{equation}
The action is invariant under the gauge transformations 
\begin{equation}    A_\mu \rightarrow A_\mu - \partial_\mu \phi  \end{equation}
\begin{equation}    B_\mu \rightarrow B_\mu - \partial_\mu \psi \end{equation}
\begin{equation}  \Psi \rightarrow e^{+ie\phi + if\gamma^5 \psi} \; \Psi .\end{equation} 
But the functional measure
\begin{equation}    \int {\cal D}(A_\mu B_\mu \bar{\Psi} \Psi)e^{i\int {\cal L}d^4x} \end{equation} 
is not, due to transformation properties of the Fermion determinant.  In particular under the gauge transformation (1.4 - 1.6) the measure in (1.7) gets multiplied by a factor 
\begin{equation} 
e^{{-i \over 8\pi^2} \int(fe^2\psi \tilde{F}^{\mu \nu}\; F_{\mu \nu} + {1 \over 3} \; f^3 \; \psi \tilde{G}^{\mu \nu}G_{\mu \nu}) d^4x }
 \end{equation} 
due to the axial anomaly, [1], [2], [3].  ($\tilde{F}$ and $\tilde{G}$ are, as usual, the dual tensors.)  The lack of gauge-invariance destroys this theory.

We define herein a new theory that has the full gauge-invariance, is renormalizable, and has the same particle content as the formal theory above.  We introduce two massless scalar fields  $a(x)$ and $b(x)$, with $b(x)$ a ghost field, and consider the Lagrangian:
\vfill\eject
\begin{eqnarray}
{\cal L}^g &=& i \bar{\Psi} \gamma^\mu(\partial_\mu + ie A_\mu + if \gamma^5 B_\mu)\Psi - {1 \over 4} F^{\mu \nu} F_{\mu \nu} - {1 \over 4} \; G^{\mu \nu} 
G_{\mu \nu} \nonumber \\
&+& {1 \over 16 \pi^2} \left( fe^2 \tilde{F}^{\mu \nu} \; F_{\mu \nu} + {1 \over 3} f^3 \tilde{G}^{\mu\nu} G_{\mu \nu} \right) (a+b) \nonumber \\
&-& {\mu^2 \over 2} (\partial_\mu a + B_\mu)^2 + {\mu^2 \over 2}(\partial_\mu b + B_\mu)^2 
\end{eqnarray} 
and the gauge transformations:
\begin{eqnarray}
A_\mu &\rightarrow& A_\mu - \partial_\mu \phi \\
B_\mu &\rightarrow& B_\mu - \partial_\mu \psi \\
\Psi &\rightarrow& e^{+ie\phi + if\gamma^5 \psi} \; \Psi \\ 
 a &\rightarrow& a + \psi \\
 b &\rightarrow & b + \psi  
\end{eqnarray}
Under a gauge transformation
\begin{equation} 
{\cal L}^g \rightarrow {\cal L}^g + {1 \over 8\pi^2} \ \psi \left(fe^2 \tilde{F}^{\mu \nu} F_{\mu \nu} + {1 \over 3} f^3 \tilde{G}^{\mu \nu} G_{\mu \nu} \right)
\end{equation} 
which is just the right factor to ensure that the measure
\begin{equation}  
\int \ {\cal D}(abA_\mu B_\mu \bar{\Psi}\, \Psi)e^{i\int {\cal L}^gd^4x} 
\end{equation} 
is invariant.  This formulation of our model was shown to us by S. Coleman.  We will see later that the fields $a$ and $b$ make no net contribution in the unitarity relations, and that the theory is renormalizable.  (The theory is independent of $\mu^2$, for $\mu^2 > 0$.)  The theory of the current paper is rather similar to the abelian Wess-Zumino model (the abelian effective gauge theory with a Wess-Zumino counterterm) [4], [5], though the development is along slightly different lines.  It may be viewed as the ``minimal" modification of this Wess-Zumino model making that theory renormalizeable. (This relationship to the Wess-Zumino model is detailed in the Appendix.)  One arrives at a new mechanism to construct theories that were once rejected as possessing anomalies.  We expect many applications of the present construction.  (Nothing here, of course, prevents the $\pi^0$ from decaying into two photons, the $\pi^0$ field given by the same operator as always.) 
\vfill\eject

My approach to this model was through the study of regularizations.  In two recent papers I presented a gauge-invariant regularization of the Weyl determinant using wavelets [6], [7].  In that work cutoff gauge-invariant expressions were obtained, but the task of letting the cutoff go to infinity was left open.  The expressions were also unwieldy and not computationally effective.  The present paper assumes no knowledge of this previous work!  The wavelet constructions may be viewed as only motivating the search for gauge-invariant theories such as here presented.

In Section 2 gauge-invariant regularizations are given of both the AVV and AAA triangle diagrams.  These are simple to construct.  Why physicists have tended to reject these regularizations is also discussed.  In Section 3 these triangle regularizations are used to give a gauge-invariant definition (regularization) of the fermion determinant.  Section 4 gives three equivalent formulations of the new gauge-invariant theory (1.9 - 1.14).  It is shown that the original theory (1.1 - 1.6) may be viewed as equivalent to the new gauge-invariant theory {\bf if} in (1.7) the fermion determinant (used in constructing the measure) is taken in the gauge-invariant form from Section 3!  Section 5 deals with renormalizability of the theory and Section 6 with the particle content.

The generalization of this treatment for general models with only abelian anomalies is straightforward and obvious for Sections 2, 3, and 4.  In particular it is clear what the actions are in any of the three formulations, and what the expression is for the gauge-invariant fermion determinant.  The proof of renormalizability would have to be technically different, though the basic ingredients are the same.

We do not know if any of the ideas of this paper can be extended to say something about the non-abelian anomaly, [8].  The work in [6] and [7] suggests this may be possible.  Some of the ideas in [9 ] may be related to the current work.

When not otherwise indicated we will follow the conventions of Sterman's book, [10]. 

\vfill\eject

\setcounter{equation}{0}

\section{Gauge-Invariant Regularization of the AVV and AAA Triangle Diagrams.}

\indent

We consider the triangle amplitude in configuration space:
\begin{equation} 
F^{\alpha \beta \gamma} (x,y,z) \ \ = \ \ {\rm Tr} \left(\gamma^\alpha \gamma^5 \; S(x,y) \gamma^\beta S(y,z) \gamma^\gamma \; S(z,x) \right) \end{equation} 
where $S$ is the massless Feynman propagator.  We let $\Gamma$ be the symmetrized sum:
\begin{equation} 
\Gamma^{\alpha \beta \gamma}(x,y,z) \ = \ F^{\alpha \beta \gamma} (x,y,z) + F^{\alpha \gamma\beta}(x,z,y). 
\end{equation} 
For $x,y,z$ distinct, this is a well-defined distribution and satisfies:
\begin{eqnarray}
&{\partial \over \partial x^\alpha}&  \Gamma^{\alpha \beta \gamma} \ (x,y,z) = 0 \\
&{\partial \over \partial y^\beta}& \Gamma^{\alpha \beta \gamma} \ (x,y,z) = 0 \\
&{\partial \over \partial z^\gamma}& \ \Gamma^{\alpha \beta \gamma} \ (x,y,z) = 0. \end{eqnarray}
By a {\bf regularization} of $\Gamma$ we mean an extension of the definition of $\Gamma$ to be a distribution defined for all $x,y,z$ (including coincident points). The usual regularization, $\Gamma_r$, satisfies (2.4) and (2.5) but not (2.3), [1], [2], [3].  In fact it satisfies the {\it anomaly} equation:
\begin{equation} 
{\partial \over \partial x^\alpha} \; \Gamma^{\alpha \beta \gamma}_r \ (x,y,z)  =  \ {i \over 4 \pi^2} \; \varepsilon^{\beta i \gamma j} {\partial \over \partial y^i} {\partial \over \partial z^j} \delta(x-y) \delta(x-z) .           
\end{equation} 

We define an alternate regularization, $\Gamma_g$, that will satisfy all three symmetry equations (2.3), (2.4), and (2.5) (as well as Lorentz invariance, the other symmetry).

\begin{equation} 
\Gamma_g^{\alpha \beta \gamma} \; (x,y,z) = \Gamma^{\alpha \beta \gamma}_r \;(x,y,z)  -  {i \over 4\pi^2}\; \varepsilon^{\beta i \gamma j} {\partial \over \partial x_\alpha} {\partial \over \partial y^i} G(x,y) {{\partial} \over {\partial z^j}}  \delta (y-z). 
\end{equation} 
$G$ is the massless Feynman boson propagation
\begin{equation}    
 G(x,y) = - \; {1 \over (2\pi)^4} \int d^4 \;k \ {1 \over k^2 + i \varepsilon} \; e^{-ik \cdot (x-y)} 
\end{equation} 
satisfying
\begin{equation} 
g^{\mu \nu} \; \partial_\mu \partial_\nu \; G \ \ = \ \ \delta(x-y) .  
\end{equation} 
We note that $\Gamma_r$ and $\Gamma_g$ are both regularizations of $\Gamma$ since they differ only at places where there are coincident arguments.

\bigskip
\noindent
\underline{Comment 1.}  \ \ This regularization of the triangle amplitude that satisfies all conservation laws (symmetries) was excluded by Adler in [3] by his required property (iii) (p. 31 of [3]):

\smallskip

{\it The difference between two regularizations ``must be a polynomial in the momentum} \\ 
\indent
{\it  variables"} ...  {\it to avoid introducing  ``spurious kinematical singularities."}

\bigskip

\noindent
\underline{Comment 2.}  The {\bf first miracle} that enables our construction to work is that  the anomaly on the right side of (2.6) is gauge-invariant.  This guarantees that the added term in (2.7) does give rise to a fully gauge-invariant regularization.  If the right side of (2.6) had been say 
\begin{equation} 
{\partial \over \partial y^\beta} {\partial \over \partial z^\gamma} \ \delta(x-y) \delta(x-z) 
\end{equation} 
instead, then the construction parallel to (2.7) would not lead to a gauge-invariant regularization.  Nor would the step of then applying projection operators to ensure conservation of the vector currents work, since that would destroy the most important property of a regularization, that it agree with the original $\Gamma$ when points are distinct.

\bigskip
\noindent
\underline{Comment 3.} The {\bf second miracle} that is involved in this paper is that the ``spurious kinematical singularities" mentioned in Comment 1 do not in fact destroy the unitarity relationship, as will be seen later.  The ``ghost particle" that is represented by the propagator in (2.7) will make no contribution as an intermediate particle in the unitarity relationships.  This is similar to the role of the Fadeev-Popov ghost in unitarity.

\vfill\eject

The treatment of the AAA triangle is similar.  We define $F^{5\;\alpha \beta \gamma}$ by replacing $\gamma^\beta$ and $\gamma^\gamma$ in (2.1) by $\gamma^\beta \gamma^5$ and $\gamma^\gamma \gamma^5$ respectively.  $\Gamma^{5\; \alpha\beta\gamma}$ is defined similarly to $\Gamma^{\alpha\beta\gamma}$
\begin{equation}    
\Gamma^{5\; \alpha\beta\gamma}\; (x,y,z) = F^{5\; \alpha\beta\gamma}\; (x,y,z) + 
F^{5\; \alpha\gamma\beta}\; (x,z,y) . \end{equation} 
The amplitude $\Gamma^{5\;\alpha\beta\gamma}_g(x,y,z)$ can then be taken to be the symmetrization of $\Gamma^{\alpha\beta\gamma}_g (x,y,z)$.
\begin{equation} 
\Gamma^{5\; \alpha\beta\gamma}_g (x,y,z) = {1 \over 3} \left( \Gamma^{\alpha\beta\gamma}_g (x,y,z) +  \Gamma^{\beta\gamma\alpha}_g (y,z,x)  + \Gamma^{\gamma\alpha\beta}_g (z,x,y) \right) . \end{equation} 
 
\vfill\eject

\setcounter{equation}{0}

\section{Gauge-Invariant Definition of the Dirac Fermion Determinant in the Perturbative Regime}.

\indent
We consider the Dirac operator for a massless fermion coupled to two abelian gauge fields, one via the vector current, the other via the axial vector current.
\begin{equation}    
d = i \; \gamma^\mu \left( \partial_\mu + ie A_\mu + i\; \gamma^5 \; fB_\mu \right). \end{equation} 
We proceed to study the problem of defining a gauge-invariant expression for Det$(d)$.  Here the gauge fields are taken as classical (unquantized) and assumed to satisfy:
\begin{description}
\item[1)] The $A_\mu(x), \ B_\mu(x)$ are ``small".  So that we are in the perturbative regime.
\item[2)] The $A_\mu(x), \ B_\mu(x)$ are zero at infinity.  So that there are no infrared difficulties.
\end{description}

We will construct an expression for the determinant $\det(d)$ that satisfies the following properties:
\begin{description}
\item[P1)]  The determinant is Lorentz invariant.  (Alternatively, one could perform a Euclidean construction.)
\item[P2)]  The determinant is gauge-invariant (under gauge transformations that are the \\ identity near infinity).
\item[P3)]  The expansion for $\ell n$(Det$(d))$ differs from the ``usual expansion" by ``added terms".  The added terms are of degree $3$, associated to the triangle diagrams.
\item[P4)]  The determinant as defined is a regularization of the perturbative expression.  The added terms correspond to the terms added in equation (2.7), so that the newly defined determinant differs from the usual expression for the determinant as the difference of two regularizations.
\item[P5)]  All the added terms vanish in the Lorentz gauge.
\end{description}

Our determinant differs from the usual determinant exactly by the fact that it uses a different regularization scheme.  The construction of fully gauge-invariant regularizations of the AVV and AAA amplitudes in Section 2 is all that is needed to yield a fully gauge-invariant expression for the determinant, as all other terms in the loop expansion are gauge-invariant as they are usually defined (by the conventional regularization schemes).

We will from now on denote as ${\rm Det}^r(d)$ the expression for the determinant as computed by the usual regularization procedure, and the newly defined gauge-invariant expression as ${\rm Det}^g(d)$ .

With $G$ the massless boson Feynman propagator we find from Section 2 that
\begin{equation}
{\rm Det}^g(d) = {\rm Det}^r(d)
\cdot e^{{ i \over 8\pi^2} \int d^4x\int d^4y B^\mu(x) {\partial \over \partial x^\mu} G(x,y) \left[ fe^2 \tilde{F}^{\mu \nu}(y) F_{\mu\nu}(y) + {1 \over 3} f^3 \tilde{G}^{\mu\nu}(y) G_{\mu\nu}(y) \right] }. 
\end{equation} 

\vfill\eject

\setcounter{equation}{0}

\section{Definition of the Quantum Field Theory.}

{\bf 4.1  Formal Preliminaries and Problems with Gauge Invariance.}

We are set to define the quantum field theory representing a massless Dirac fermion coupled to two abelian gauge fields.  We introduce the Lagrangian densities
\begin{eqnarray}
{\cal L}_F   &=& \bar{\Psi} d \Psi = \bar{\Psi}\,i \gamma^\mu \left(\partial_\mu + ie A_\mu + i\, \gamma^5 f B_\mu \right)\Psi\     \\
  {\cal L}_B   &=&  \   - \; {1 \over 4} \; F_{\mu \nu} F^{\mu \nu} - \; {1 \over 4} \; G_{\mu \nu} G^{\mu \nu} \\
{\cal L}  &=&  {\cal L}_F \ \  + \ \   {\cal L}_B
\end{eqnarray}
where
\begin{eqnarray}
F_{\mu \nu} &=& \partial_\mu A_\nu - \partial_\nu A_\mu    \\
G_{\mu \nu} &=& \partial_\mu B_\nu - \partial_\nu B_\mu.    
\end{eqnarray}
We wish to be able to define expectation values via the functional integral:
\begin{equation}   
< {\it p} > \ = \ {\cal N} \int {\cal D} (A_\mu B_\mu \bar{\Psi} \Psi)e^{i\;S} \ {\it p} 
\end{equation} 
with 
\begin{equation} S \ \ = \ \ \int \ d^4 x \ {\cal L}.  \end{equation} 

We let ${\it p}$ be a polynomial in $\bar{\Psi}, \Psi$ of the form
\begin{equation} 
 {\it p} \ = \ T  \Psi_{\alpha_n}(y_n) \cdots \Psi_{\alpha_1}(y_1) 
\bar{\Psi}_{\beta_1}(x_1) \cdots \bar{\Psi}_{\beta_n}(x_n) q(A_\mu, B_\mu).
\end{equation} 
(It is clearly sufficient to be able to evaluate (4.6) for expressions of this form.)  Performing the fermionic integrals in (4.6) we arrive at the form
$$
\left< {\it p} \right> \ = \ {\cal N} \int {\cal D} (A_\mu B_\mu) e^{iS_B} \cdot \ {\rm Det}^r(d) \cdot q(A_\mu, B_\mu) \cdot
$$
\begin{equation} 
\cdot \ i^n \ \cdot \   \sum_P (-1)^{{\rm Sign}(P)} \prod^n_{\kappa = 1} S^b_{\alpha_{P(\kappa)}\beta_\kappa} \left( y_{P(\kappa)}, x_\kappa, A_\mu, B_\mu \right)	
\end{equation} 
where the sum is over permutations of $1, ... , n$ and
\begin{equation}    S_B \ \ = \ \ \int \ d^4 x \; {\cal L}_B .  \end{equation} 
$S^b$ is the Feynman fermion propagator in the background $A_\mu, B_\mu$ field, or equivalently
\begin{equation}    S^b \ \ = \ \ d^{-1}     \end{equation} 
with time-ordered choice of boundary conditions.  Note that the usual regularization of the determinant is employed in (4.9).

The Lagrangian ${\cal L}$ in (4.3) is invariant under the gauge transformation (defined by  $\phi$ and $\psi$):
\begin{eqnarray}
A_\mu &\rightarrow& A_\mu - \partial_\mu \phi \\
B_\mu &\rightarrow& B_\mu - \partial_\mu \psi \\
\Psi &\rightarrow& e^{+ie\phi + if\psi \gamma^5} \; \Psi \ .
\end{eqnarray}
However, when we refer to equation (4.9) which really defines how computations will be performed the expression would transform correctly except that under a gauge change
\begin{equation}
{\rm Det}^r(d) \rightarrow e^{{ -i \over 8\pi^2} \int \left(fe^2 \psi \tilde{F}^{\mu\nu} F_{\mu\nu} + {1 \over 3} \; f^3 \psi \tilde{G}^{\mu\nu} G_{\mu\nu}  \right)d^4x} \; 
{\rm Det}^r(d).
\end{equation}
This follows from equation (3.2); or standard computation.  The determinant would have to be invariant for (4.9) to transform correctly (set $ {\it p}  = 1$ in (4.9) to see this).

One needs gauge-invariance to get a satisfactory theory, to insure renormalizability, 
to obtain the correct degrees of freedom for the $B_\mu$ field, and to enable some sort of gauge fixing  to make the functional integral meaningful.  We modify the above formal procedure to obtain a satisfactory quantum field theory, one that is renormalizable, maintains all the gauge-invariance, and has the correct particle structure.  We present three \underline{equivalent} formulations of this new theory.

\vfill\eject

{\bf 4.2.  The New Theory, Formulation 1, Gauge-Invariant Regularization.}

In this formulation we modify the formal theory of Subsection 4.1 by using ${\rm Det}^g(d)$ in (4.9), instead of ${\rm Det}^r(d)$.  This determinant defined in Sections 2 and 3 is gauge-invariant.  One may use a standard gauge fixing, and renormalization scheme, to construct the theory from (4.9) with the determinant replacement.

Were it not for the historical development of the subject, it would be just as natural to have used ${\rm Det}^g(d)$  instead of ${\rm Det}^r(d)$ right at the outset.  We feel the regularization of Section 2 is preferable to the usual regularization.

{\bf 4.3.  The New Theory, Formulation 2, Via a Nonlocal Action.}

Referring to equation (3.2) we see that working with ${\rm Det}^g(d)$  instead of ${\rm Det}^r(d)$  as in the last subsection is equivalent to introducing the exponential factor in (3.2) into the action.  Thus in the second formulation we construct our new theory with the functional measure:
\begin{equation}
\int {\cal D} (A_\mu B_\mu \bar{\Psi} \Psi) \ e^{iS^{NL}}
\end{equation}
$$
S^{NL} = \int d^4 x \left\{  i\bar{\Psi} \gamma^\mu (\partial_\mu + ieA_\mu + if\gamma^5 B_\mu)\Psi -  {1 \over 4 } F_{\mu\nu} F^{\mu\nu} - \; {1 \over 4} G_{\mu\nu} G^{\mu\nu} \right\}
$$
\begin{equation}
+ \; {1 \over 8\pi^2} \int d^4 x \int d^4 y B^\mu(x) {\partial \over \partial x^\mu} G(x,y) \left[ fe^2 \tilde{F}^{\mu\nu}(y)F_{\mu\nu}(y) + {1 \over 3} \; f^3 \; \tilde{G}^{\mu\nu}(y) G_{\mu\nu}(y) \right]  .
\end{equation}
$F_{\mu\nu}$ and $G_{\mu\nu}$ are defined as in (4.4) and (4.5).  This measure is invariant under the gauge transformations (4.12 - 4.14), where the fermion determinant is treated as usual.

This action is nonlocal, as would be the action of a non-abelian gauge field in the presence of the Fadeev-Popov determinant before ghost fields are introduced.  The propagator $G$ will be seen not to destroy unitarity (and has the right $i \varepsilon$ prescription not to destroy causality either).  That unitarity is not destroyed is the {\bf second miracle} we mentioned before in Section 2.

\vfill\eject

{\bf 4.4.  The New Theory, Formulation 3, Using a Ghost Field.}

Our final formulation of the new theory is the one given in the introduction (1.9 - 1.14).  One easily sees that integrating out the $a$ and $b$ fields converts formulation 3 to formulation 2.

\vfill\eject

\setcounter{equation}{0}

\section{Perturbation Theory and Renormalization.}

Our development of perturbation theory closely follows the classic work of Adler and Bardeen in [11] and [12].  We address the infrared problem by giving the $A_\mu$ and $B_\mu$ mesons a (small) mass, $m$.  These mesons are unstable, and the precise meaning of their mass, $m$, is specified below.  Referring to Section 4 the point of view we take towards our model in this section is closest to that of Formalism 1.  We now present our ``regulated" Action ${\cal S}^R$ (we will be seen to follow closely reference [11], even as to notation):
\begin{eqnarray}
 {\cal S}^R &=& \int \Big\{ i \bar{\Psi} \gamma^\mu \left( \partial_\mu + ie_0 (A_\mu + A^R_\mu) + if_0\gamma^5(B_\mu + B^R_\mu ) \right) \Psi  \nonumber \\
&-& {1 \over 4} \; F_{\mu\nu} F^{\mu\nu} \ + \ {1 \over 2} \; m^2_{0A} A_\mu A^\mu \nonumber  \\
&-& {1 \over 4} \; G_{\mu\nu} G^{\mu\nu} \ + \ {1 \over 2} \; m^2_{0B} B_\mu B^\mu \\
&+& {1 \over 4} \; F^R_{\mu\nu} F^{R\mu\nu} \ - \ {1 \over 2} \; \Lambda^2 \; A^R_\mu \; A^{R\mu} \nonumber \\
&+&  {1 \over 4} \; G^R_{\mu\nu} G^{R\mu\nu} \ - \ {1 \over 2} \; \Lambda^2 \; B^R_\mu \; B^{R\mu}  \Big\} \; d^4x + \ {\rm C.F.}(2) + {\rm C.F.(3)} . \nonumber
\end{eqnarray}
$A^R_\mu$ and $B^R_\mu$ are the ghost regulator fields.  $\Lambda$ is the regulator (cutoff) mass.  The bare parameters $e_0,\  f_0, \  m_{0A}, \  m_{0B}$ are each functions of the physical parameters $e, f, m$  and  $\Lambda$.  C.F.(2), which also appears in [11], is an explicit counterterm
\begin{equation}
{\rm C.F.}(2) = \ {\rm C.F.}(2)_A + \ {\rm C.F.}(2)_B
\end{equation}
where
\begin{equation}
{\rm C.F.}(2)_A = c_{2A} \int \left(F_{\mu\nu} + F_{\mu\nu}^R \right) \left( F^{\mu\nu}
+ F^{R\mu\nu} \right) d^4 x
\end{equation}
\begin{equation}
{\rm C.F.}(2)_B = c_{2B} \int \left(G_{\mu\nu} + G_{\mu\nu}^R \right) \left( G^{\mu\nu}
+ R^{R\mu\nu} \right) d^4 x
\end{equation}
and each $c_2$ is chosen so that the two vertex vacuum polarization loops satisfy
\begin{equation}
\Pi^{(2)}_{\mu\nu} (q) = (q_\mu q_\nu - g_{\mu\nu}\, q^2 ) \Pi^{(2)} (q^2)
\end{equation}
with
\begin{equation}
{\it Re} \left( \Pi^{(2)} (m^2) \right) = 0 .
\end{equation}
(We have not indicated the $A$ and $B$ subscripts here.  We adhere to standard conventions that ensure that the two and four vertex loops automatically preserve vector and axial vector current conservation.)  C.F.(3) is the key new term ensuring that the three vertex loop (with the ``counterterm" C.F.(3) added to it) satisfies both current conservations.  We find from (4.17) that
$$
{\rm C.F. (3)} \ =  \ {1 \over 8\pi^2} \int d^4 x \int d^4 y \left( B^\mu(x) + B^{R\mu}(x) \right) {\partial \over \partial x^\mu} G(x,y) . 
$$
\begin{equation}
\left[ f_0 e_0^2 \left(\tilde{F}^{\mu\nu}(y) + \tilde{F}^{R\mu\nu}(y) \right) \left(F_{\mu\nu} (y)
+ F^R_{\mu\nu}(y)  \right)  \right.
\end{equation}
$$
\left. + {1 \over 3} \; f^3_0 \left( \tilde{G}^{\mu\nu}(y) + \tilde{G}^{R\mu\nu}(y) \right)
\left(G_{\mu\nu} (y) + G^R_{\mu\nu}(y) \right)  \right] 
$$

With the two and three vertex fermion loops handled by hand above, all integrals of perturbation terms in the regulated theory are finite, and algebraic manipulations, such as leading to the Ward identities, may be done by hand.  Slightly generalizing  the identities in [11] to a model with zero fermion mass we get the identities, Ward and commutator types:
\begin{eqnarray}
\gamma^5 S'_F &=& - \; S'_F \gamma^5 \\
\gamma^5 \Gamma_\mu(p,p')  &=& - \Gamma_\mu (p,p')\gamma^5 \\
\gamma^5 \Gamma^5_\mu(p,p')  &=& - \Gamma^5_\mu (p,p')\gamma^5 \\
(p-p') \Gamma_\mu (p,p') &=& S'_F(p)^{-1} - S'_F(p')^{-1} \\
(p-p') \Gamma^5_\mu (p,p') &=& S'_F(p)^{-1} \gamma^5 - S'_F(p')^{-1} \gamma^5 .
\end{eqnarray}
Each of these identities holds in the regulated theory, with the vertex parts defined similarly to in     \ Q.E.D.

Renormalization constants are defined similar to in \ Q.E.D.

\vfill\eject

\begin{eqnarray}
Z^{-1}_2 \; S'_F(p) &=& \tilde{S}'_F(p) \\
Z^{-1}_{3A} \; D'_{FA}(q)_{\mu \nu} &=& \tilde{D}'_{FA}(q)_{\mu\nu} \\ 
Z^{-1}_{3B} \; D'_{FB}(q)_{\mu \nu} &=& \tilde{D}'_{FB}(q)_{\mu\nu} \\
Z_1\Gamma_\mu(p,p') &=& \tilde{\Gamma}_\mu(p,p') \\
Z_1 &=& Z_2 \\
e_0 &=& Z^{-{1\over 2}}_{3A} \ e \\
Z_1\Gamma^5_\mu(p,p') &=& \tilde{\Gamma}^5_\mu(p,p') \\
f_0 &=& Z^{-{1 \over 2}}_{3B} \ f
\end{eqnarray}
with the usual wave function renormalizations on external lines.  The values of $Z_2, Z_{3A}, Z_{3B}$, $m_{0A}$ and $m_{0B}$ are determined by requiring 
\begin{equation}
\lim_{p \rightarrow p_0} \ p_\mu \gamma^\mu \tilde{S}'_F(p) u(p_0) \ = \ u(p_0)
\end{equation}
if $p_0$ is not identically zero and
\begin{eqnarray}
p^2_0 \ &=& \ 0 \\
p_{0\mu} \gamma^\mu \; u(p_0) &=& 0 ,
\end{eqnarray}
and
\begin{equation}
\lim_{p \rightarrow p_0} (-p^2 + m^2) \hat{D}'_{FA}(p)_{\mu\nu}\; n^\nu(p_0) = n_\mu(p_0)
\end{equation}
if
\begin{eqnarray}
p^2_0 &=& m^2 \\
p^\mu_0 \; n_\mu (p_0) &=& 0,  
\end{eqnarray}
where
\begin{equation}
\hat{D}'_{FA} = \left( {\it Re} (\tilde{D}'_{FA})^{-1} \right)^{-1} ,
\end{equation}
and a similar set of equations for $\tilde{D}'_{FB}$.  The renormalization procedure is now standard.  We do not discuss the problem of taking the limit of $m$ to zero.

\vfill\eject

\setcounter{equation}{0}

\section{Particle Structure}

We look at the unitarity relationship for our theory via the naive cut relationships for the 
Feynman diagrams generated.  We think in terms of Formalism 2 of Section 4.  The propagator  $G(x,y)$ in the nonlocal term in the action is a Feynman propagator so the analytic structure generated is the same as for usual local Lagrangians.  Only transverse terms in the polarizations of the $A_\mu$ and $B_\mu$ vector fields contribute in intermediate states of the unitarity relationship.  This is by the usual argument (see for example p. 347, [10]).  Other states of the bosons make no contribution, using the fact that the boson is coupled to conserved currents (in loops or external on-shell lines).  The same argument shows that the propagator $G$, corresponding to a ghost particle, makes no net contribution when it appears in intermediate states.  Viewing the expression
\begin{equation}
\int d^4 x \ B^\mu(x) \ {\partial \over \partial x^\mu} \; G(x,y) \dots 
\end{equation}
from (4.17) we see that after integration by parts the propagator $G$ connects to ${\partial B^\mu \over \partial x^\mu}$.  Thus again we can use that this ${\partial B^\mu \over \partial x^\mu}$  couples to a conserved current in the $S$ matrix element and the net contribution vanishes.  (In both cases we also need the fact that bosons in external lines are all transverse, so that if ${\partial B^\mu \over \partial x^\mu}$ hits an external boson line it vanishes.  One is showing that the $S$ matrix, restricted to fermion and transverse $A_\mu \; , \; B_\mu$ states, is unitary.)  If we had used Formalism 1 with the two auxiliary boson fields $a(x)$ and $b(x)$ of (1.9), we would have discovered that the contributions of the $a$ and $b$ fields cancel each other.

\vfill\eject

\noindent
{\SC APPENDIX.  The (abelian)  Wess-Zumino Model .}

For the Wess-Zumino Lagrangian we take
$$
{\cal L}^{WZ} = i \bar{\Psi} \gamma^\mu(\partial_\mu + ie A_\mu + i \gamma^5 f B_\mu)\Psi - {1 \over 4} F^{\mu \nu} F_{\mu \nu} - {1 \over 4} \; G^{\mu \nu} 
G_{\mu \nu}$$
$$ + {1 \over 8 \pi^2} \left( fe^2 \tilde{F}^{\mu \nu} F_{\mu \nu}  + {1 \over 3} f^3 \tilde{G}^{\mu\nu}G_{\mu\nu} \right) a $$
$$- {\mu^2 \over 2} (\partial_\mu a + B_\mu)^2 \eqno(A.1)  $$

\noindent
Under the gauge transformations:

$$  A_\mu \rightarrow A_\mu - \partial_\mu \phi  \eqno(A.2)$$
$$  B_\mu \rightarrow B_\mu - \partial_\mu \psi \eqno(A.3)$$
$$ \ \ \ \   \Psi \rightarrow e^{+ie\phi + if\gamma^5 \psi} \; \Psi \eqno(A.4)$$ 
$$  a \rightarrow a + \psi  \eqno(A.5)$$
the measure
$$    \int {\cal D}\big( aA_\mu B_\mu \bar{\Psi} \Psi \big)e^{i\int {\cal L}^{WZ}d^4x} \eqno(A.6) $$
is invariant.  If we integrate out the $a$ field we construct a theory with the functional measure
$$	\int {\cal D} \big(A_\mu B_\mu \bar{\Psi}\Psi \big)e^{iS^{WZ}}
\eqno(A.7)   $$
where
$$
S^{WZ} = S^{NL} - {1 \over 2\mu^2} \int d^4x \int d^4y s(x) G(x,y) s(y) \eqno(A.8)
$$
with
$$ s(x) = {1 \over 8 \pi^2} \left( fe^2 \tilde{F}^{\mu \nu}(x) \; F_{\mu \nu}(x) + {1 \over 3} f^3 \tilde{G}^{\mu\nu}(x) G_{\mu \nu} (x) \right) \eqno(A.9) $$
$G(x,y)$ is defined in (2.8), and $S^{NL}$ is given in (4.17).  The second term on the right side of (A.8) leads to the non-renormalizability of the Wess-Zumino model.  Our model is obtained from the Wess-Zumino model by throwing away this {\bf gauge-invariant} term.
\vfill\eject

\centerline{{\bf REFERENCES}}

\begin{description}
\item[[1]]  S.L. Adler, ``Axial-Vector Vertex in Spinor Electrodynamics'', {\it Phys. Rev.} {\bf 177}, 2426 (1969).
\item[[2]]  J.S. Bell and R. Jackiw, `` A PCAC Puzzle: $\pi^0 \rightarrow \gamma \gamma$ in the Sigma Model'', Nuovo Cim. {\bf 60A}, 47 (1969).
\item[[3]]  S.L. Adler, ``Perturbation Theory Anomalies'', in \underline{Lectures in Elementary Particle Physics}, ed. S. Deser, M. Grisaru and H. Pendleton (M.I.T. Press, Cambridge, MA.) (1970).
\item[[4]]  J. Wess and B. Zumino, ``Consequences of Anomalous Ward Identities", {\it Phys. Lett. } {\bf B37}, 95 (1971).
\item[[5]]  J. Preskill, ``Gauge Anomalies in an Effective Field Theory", {\it Ann. Phys.} {\bf 210}, 323 (1991).
\item[[6]]  P. Federbush, ``A New Formulation and Regularization of Gauge Theories Using a Non-Linear Wavelet Expansion'',  {\it Prog. Theor. Phys.},  {\bf 94}, No. 1135 (1995) (on-line hep-ph/9505368).
\item[[7]]  P. Federbush, ``A Gauge-Invariant Regularization of the Weyl Determinant Using Wavelets'', preprint (on-line hep-th/9509131).
\item[[8]]   W.A. Bardeen, ``Anomalous Ward Identities in Spinor Field Theories'', {\it Phys. Rev.} {\bf 184}, 1848 (1969).
\item[[9]]  T. D. Kieu, ``Chiral Gauge Theory in Four Dimensions", preprint. 
\item[[10]]  G. Sterman, \underline{An Introduction to Quantum Field Theory}, Cambridge University Press, (1993).
\item[[11]]  S.L. Adler and W.A. Bardeen, ``Absence of Higher Order Corrections in the Anomalous Axial-Vector Divergence Equations", {\it Phys. Rev.} {\bf 182} , 1517 (1969).
\item[[12]]  W.A. Bardeen, ``Regularization of Gauge Field Theories", In {\it Proc. of the XVI Intern. Conf. on High Energy Physics, NAL.}, eds. Jackson, J.D. \& Roberts, A. , Fermilab, Batavia (1972). 
\end{description}

\end{document}